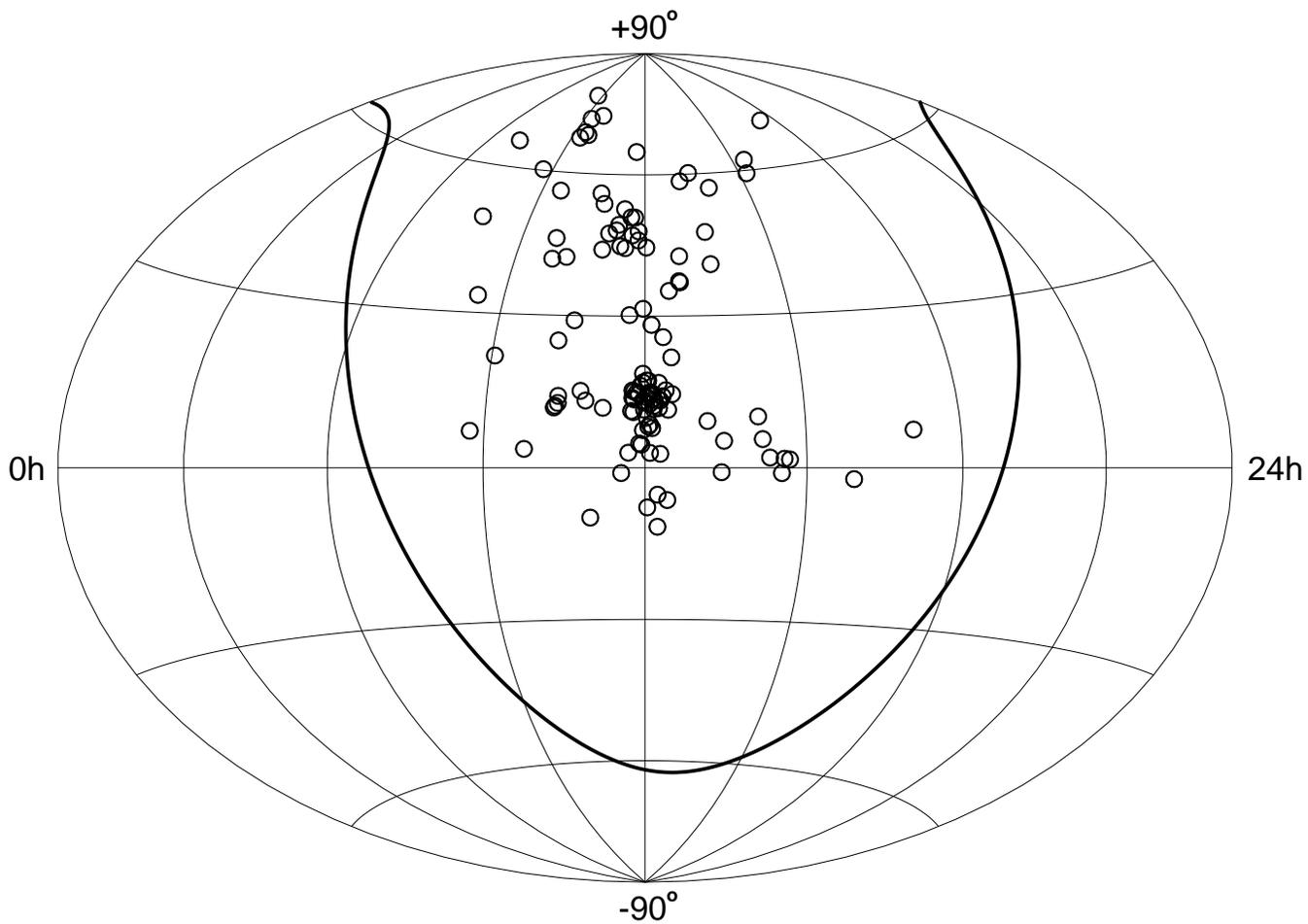

Figure 1.

TABLE 1. Properties of galaxies in the catalog.

| NGC | Obs. | $T$ | $\alpha$ (1950) | $\delta$ (1950) | $B_T$ | $\langle v \rangle_{hc}$ | $D_{25}$ | $I$ | p.a. |
|---|---|---|---|---|---|---|---|---|---|
| 2403 | Palomar | 6 | 07:32:02.7 | +65:42:43 | 8.93 | 129 | 1312 | 55.8 | 127 |
| 2541 | Palomar | 6 | 08:11:01.9 | +49:13:00 | 12.26 | 556 | 378 | 59.9 | 165 |
| 2683 | Lowell | 3 | 08:51:54.2 | +33:27:34 | 10.64 | 402 | 559 | 76.4 | 44 |
| 2715 | Lowell | 5 | 09:06:01.0 | +78:06:17 | 11.79 | 1317 | 293 | 70.2 | 22 |
| 2768 | Lowell | −5 | 09:07:45.2 | +60:14:40 | 10.84 | 1334 | 487 | 58.3 | 95 |
| 2775 | Lowell | 2 | 09:09:46.1 | +07:05:06 | 11.03 | 1350 | 255 | 39.1 | 155 |
| 2903 | Palomar | 4 | 09:29:19.9 | +21:43:11 | 9.68 | 556 | 755 | 61.4 | 17 |
| 2976 | Lowell | 5 | 09:45:58.5 | +67:57:18 | 10.82 | 1 | 353 | 62.8 | 143 |
| 2985 | Lowell | 2 | 09:45:52.6 | +72:30:45 | 11.18 | 1317 | 274 | 37.4 | 0 |
| 3031 | Palomar | 2 | 09:51:29.9 | +69:18:19 | 7.89 | −37 | 1614 | 58.3 | 157 |
| 3077 | Lowell | 12 | 10:02:09.1 | +68:43:11 | 10.61 | 11 | 322 | 33.7 | 45 |
| 3079 | Lowell | 5 | 09:58:35.4 | +55:55:11 | 11.54 | 1124 | 476 | 79.5 | 165 |
| 3147 | Lowell | 4 | 10:15:23.4 | +73:27:23 | 11.43 | 2810 | 233 | 27.0 | 155 |
| 3166 | Lowell | 0 | 10:11:09.3 | +03:40:25 | 11.32 | 1344 | 287 | 60.7 | 87 |
| 3184 | Lowell | 6 | 10:17:38.7 | +41:28:29 | 10.36 | 589 | 444 | 21.1 | 135 |
| 3198 | Palomar | 5 | 10:16:53.0 | +45:47:59 | 10.87 | 663 | 510 | 67.1 | 35 |
| 3319 | Palomar | 6 | 10:36:14.1 | +41:56:46 | 11.48 | 745 | 369 | 56.7 | 37 |
| 3344 | Lowell | 4 | 10:42:48.5 | +24:58:46 | 10.45 | 584 | 424 | 24.2 | 0 |
| 3351 | Lowell | 3 | 10:43:15.6 | +11:45:15 | 10.53 | 777 | 444 | 47.5 | 13 |
| 3368 | Lowell | 2 | 10:44:06.9 | +12:05:05 | 10.11 | 898 | 455 | 46.2 | 5 |
| 3377 | Lowell | −5 | 10:47:09.1 | +14:02:35 | 11.24 | 691 | 314 | 54.9 | 35 |
| 3379 | Lowell | −5 | 10:47:12.6 | +12:38:22 | 10.24 | 888 | 322 | 27.0 | 0 |
| 3486 | Lowell | 5 | 10:59:44.8 | +29:02:10 | 11.05 | 680 | 424 | 42.2 | 80 |
| 3556 | Lowell | 6 | 11:10:40.1 | +55:43:56 | 10.69 | 693 | 522 | 75.1 | 80 |
| 3596 | Lowell | 5 | 11:12:27.9 | +15:03:38 | 11.95 | 1191 | 238 | 17.3 | 0 |
| 3623 | Lowell | 1 | 11:18:19.5 | +13:09:31 | 10.25 | 806 | 586 | 72.8 | 174 |
| 3631 | Lowell | 5 | 11:18:13.3 | +53:26:43 | 11.01 | 1157 | 300 | 17.3 | 0 |
| 3672 | Lowell | 5 | 11:24:31.1 | −09:44:33 | 12.09 | 1862 | 250 | 62.1 | 12 |
| 3675 | Lowell | 3 | 11:25:33.3 | +43:38:45 | 11.00 | 765 | 353 | 58.3 | 178 |
| 3726 | Lowell | 5 | 11:32:39.1 | +47:05:49 | 10.91 | 848 | 369 | 46.2 | 10 |
| 3810 | Lowell | 5 | 11:38:23.5 | +11:44:55 | 11.35 | 993 | 255 | 44.9 | 15 |
| 3877 | Lowell | 5 | 11:43:29.4 | +47:46:18 | 11.79 | 902 | 329 | 76.4 | 35 |
| 3893 | Lowell | 5 | 11:46:01.1 | +48:59:20 | 11.16 | 971 | 268 | 51.9 | 165 |
| 3938 | Lowell | 5 | 11:50:13.6 | +44:24:07 | 10.90 | 807 | 322 | 24.2 | 0 |
| 3953 | Lowell | 4 | 11:53:02.2 | +52:22:19 | 10.84 | 1053 | 415 | 59.9 | 13 |
| 4013 | Lowell | 3 | 11:57:51.8 | +44:01:12 | 12.19 | 836 | 314 | 78.8 | 66 |
| 4030 | Lowell | 4 | 11:59:45.8 | −01:02:30 | 11.42 | 1460 | 250 | 43.6 | 27 |
| 4088 | Lowell | 4 | 12:04:54.3 | +50:35:47 | 11.15 | 758 | 345 | 67.1 | 43 |

TABLE 1. (Continued.)

| NGC | Obs. | $T$ | $\alpha$ (1950) | $\delta$ (1950) | $B_T$ | $\langle v \rangle_{hc}$ | $D_{25}$ | $I$ | p.a. |
|---|---|---|---|---|---|---|---|---|---|
| 4123 | Lowell | 5 | 12:07:30.9 | +02:57:17 | 11.98 | 1328 | 261 | 42.2 | 135 |
| 4125 | Lowell | −5 | 12:07:25.1 | +65:13:41 | 10.65 | 1354 | 345 | 56.7 | 95 |
| 4136 | Lowell | 5 | 12:06:45.7 | +30:12:18 | 11.69 | 607 | 238 | 21.1 | 0 |
| 4144 | Lowell | 6 | 12:07:28.3 | +46:44:07 | 12.05 | 267 | 361 | 77.4 | 104 |
| 4157 | Lowell | 5 | 12:10:26.7 | +50:33:20 | 12.66 | 2294 | 134 | 39.1 | 115 |
| 4178 | Palomar | 8 | 12:10:13.8 | +11:08:48 | 11.90 | 376 | 307 | 69.2 | 30 |
| 4189 | Palomar | 6 | 12:11:13.9 | +13:42:11 | 12.51 | 2112 | 143 | 43.6 | 85 |
| 4192 | Palomar | 2 | 12:11:15.9 | +15:10:49 | 10.95 | −142 | 586 | 73.6 | 155 |
| 4216 | Palomar | 3 | 12:13:20.9 | +13:25:22 | 10.99 | 129 | 487 | 77.4 | 19 |
| 4242 | Lowell | 8 | 12:16:57.5 | +45:40:37 | 11.37 | 517 | 300 | 40.7 | 25 |
| 4254 | Palomar | 5 | 12:16:18.0 | +14:41:42 | 10.44 | 2407 | 322 | 29.4 | 0 |
| 4258 | Palomar | 4 | 12:16:29.0 | +47:35:00 | 9.10 | 450 | 1117 | 67.1 | 150 |
| 4303 | Palomar | 4 | 12:19:22.0 | +04:45:04 | 10.18 | 1569 | 387 | 27.0 | 0 |
| 4321 | Palomar | 4 | 12:20:22.9 | +16:06:01 | 10.05 | 1585 | 444 | 31.7 | 30 |
| 4340 | Lowell | −1 | 12:23:04.9 | +16:46:54 | 12.10 | 915 | 212 | 37.4 | 102 |
| 4365 | Lowell | −5 | 12:23:56.0 | +07:22:41 | 10.52 | 1227 | 415 | 43.6 | 40 |
| 4374 | Lowell | −5 | 12:24:30.1 | +12:56:41 | 10.09 | 956 | 387 | 29.4 | 135 |
| 4394 | Palomar | 3 | 12:23:24.9 | +18:29:23 | 11.73 | 920 | 217 | 27.0 | 0 |
| 4406 | Lowell | −5 | 12:23:39.7 | +13:13:25 | 9.83 | −248 | 534 | 49.8 | 130 |
| 4414 | Palomar | 5 | 12:23:56.9 | +31:29:55 | 10.96 | 718 | 217 | 55.8 | 155 |
| 4429 | Lowell | −1 | 12:24:54.2 | +11:23:05 | 11.02 | 1137 | 337 | 62.8 | 99 |
| 4442 | Lowell | −2 | 12:27:32.3 | +09:50:48 | 11.38 | 530 | 274 | 67.1 | 87 |
| 4449 | Lowell | 10 | 12:27:36.0 | +44:09:15 | 9.99 | 201 | 369 | 44.9 | 45 |
| 4450 | Lowell | 2 | 12:27:58.3 | +17:08:37 | 10.90 | 1957 | 314 | 42.2 | 175 |
| 4472 | Lowell | −5 | 12:29:13.8 | +08:03:42 | 9.37 | 915 | 613 | 35.6 | 155 |
| 4477 | Lowell | −3 | 12:29:29.4 | +13:41:46 | 11.38 | 1348 | 228 | 24.2 | 15 |
| 4486 | Lowell | −4 | 12:30:17.4 | +12:27:00 | 9.59 | 1282 | 499 | 37.4 | 0 |
| 4487 | Lowell | 6 | 12:28:29.5 | −07:46:41 | 11.63 | 1037 | 250 | 47.5 | 75 |
| 4498 | Palomar | 7 | 12:29:07.9 | +17:07:49 | 12.79 | 1505 | 177 | 57.5 | 133 |
| 4501 | Palomar | 3 | 12:29:27.9 | +14:41:43 | 10.36 | 2278 | 415 | 57.5 | 140 |
| 4526 | Lowell | −2 | 12:33:34.0 | +07:45:58 | 10.66 | 463 | 434 | 70.7 | 113 |
| 4527 | Palomar | 4 | 12:31:35.0 | +02:55:42 | 11.38 | 1734 | 369 | 70.2 | 67 |
| 4535 | Palomar | 5 | 12:31:47.9 | +08:28:35 | 10.59 | 1957 | 424 | 44.9 | 0 |
| 4548 | Palomar | 3 | 12:32:55.1 | +14:46:22 | 10.96 | 485 | 322 | 37.4 | 150 |
| 4559 | Palomar | 6 | 12:33:29.0 | +28:14:07 | 10.46 | 814 | 642 | 66.0 | 150 |
| 4564 | Lowell | −5 | 12:35:51.2 | +11:30:17 | 12.05 | 1119 | 212 | 65.4 | 47 |
| 4569 | Palomar | 2 | 12:34:17.9 | +13:26:25 | 10.26 | −237 | 572 | 62.8 | 23 |
| 4571 | Palomar | 7 | 12:34:25.0 | +14:29:34 | 11.82 | 341 | 217 | 27.0 | 55 |

TABLE 1. (Continued.)

| NGC | Obs. | $T$ | $\alpha$ (1950) | $\delta$ (1950) | $B_T$ | $\langle v \rangle_{hc}$ | $D_{25}$ | $I$ | p.a. |
|---|---|---|---|---|---|---|---|---|---|
| 4579 | Palomar | 3 | 12:35:12.0 | +12:05:34 | 10.48 | 1521 | 353 | 37.4 | 95 |
| 4593 | Lowell | 3 | 12:39:01.4 | −05:16:29 | 11.67 | 2497 | 233 | 42.2 | 0 |
| 4594 | Lowell | 1 | 12:39:29.9 | −11:33:57 | 8.98 | 1090 | 522 | 66.0 | 90 |
| 4621 | Lowell | −5 | 12:41:26.2 | +11:42:53 | 10.57 | 431 | 322 | 46.2 | 165 |
| 4636 | Lowell | −5 | 12:42:12.8 | +02:44:48 | 10.43 | 1018 | 361 | 39.1 | 150 |
| 4651 | Palomar | 5 | 12:41:13.0 | +16:40:05 | 11.39 | 805 | 238 | 48.6 | 80 |
| 4654 | Palomar | 6 | 12:41:25.9 | +13:23:59 | 11.10 | 1035 | 293 | 54.9 | 128 |
| 4689 | Palomar | 4 | 12:45:14.9 | +14:02:04 | 11.60 | 1617 | 255 | 35.6 | 0 |
| 4710 | Lowell | −1 | 12:49:08.5 | +15:13:27 | 11.91 | 1119 | 293 | 76.1 | 27 |
| 4725 | Palomar | 2 | 12:47:59.9 | +25:46:30 | 10.11 | 1206 | 642 | 44.9 | 35 |
| 4731 | Lowell | 6 | 12:50:30.9 | −06:19:11 | 11.90 | 1495 | 396 | 60.7 | 95 |
| 4754 | Lowell | −3 | 12:51:49.8 | +11:22:12 | 11.52 | 1396 | 274 | 57.5 | 23 |
| 4826 | Lowell | 2 | 12:56:08.7 | +21:43:50 | 9.36 | 412 | 600 | 57.5 | 115 |
| 4861 | Lowell | 9 | 12:56:40.3 | +35:07:56 | 12.90 | 843 | 238 | 68.2 | 15 |
| 4866 | Lowell | −1 | 12:56:57.9 | +14:26:25 | 12.14 | 1987 | 378 | 77.7 | 87 |
| 5005 | Lowell | 4 | 13:10:25.8 | +37:06:53 | 10.61 | 948 | 345 | 61.4 | 65 |
| 5033 | Palomar | 5 | 13:11:07.9 | +36:51:46 | 10.75 | 877 | 642 | 62.1 | 170 |
| 5055 | Palomar | 4 | 13:13:34.8 | +42:17:48 | 9.31 | 504 | 755 | 54.9 | 105 |
| 5204 | Lowell | 9 | 13:29:08.1 | +58:27:20 | 11.73 | 202 | 300 | 52.9 | 5 |
| 5248 | Lowell | 4 | 13:35:32.4 | +09:08:23 | 10.97 | 1153 | 369 | 43.6 | 110 |
| 5322 | Lowell | −5 | 13:48:36.0 | +60:14:50 | 11.14 | 1915 | 353 | 48.6 | 95 |
| 5334 | Lowell | 5 | 13:50:20.3 | −00:52:05 | 11.99 | 1382 | 250 | 43.6 | 15 |
| 5364 | Lowell | 4 | 13:53:41.1 | +05:15:33 | 11.17 | 1240 | 405 | 49.8 | 30 |
| 5371 | Lowell | 4 | 13:55:04.4 | +40:30:53 | 11.32 | 2554 | 261 | 37.4 | 8 |
| 5377 | Lowell | 1 | 13:55:46.3 | +47:17:24 | 12.24 | 1793 | 222 | 55.8 | 20 |
| 5585 | Lowell | 7 | 14:19:26.8 | +56:46:43 | 11.20 | 305 | 345 | 49.8 | 30 |
| 5669 | Lowell | 6 | 14:32:13.8 | +09:57:12 | 12.03 | 1374 | 238 | 44.9 | 50 |
| 5701 | Lowell | 0 | 14:36:41.5 | +05:34:50 | 11.76 | 1506 | 255 | 17.3 | 0 |
| 5746 | Lowell | 3 | 14:44:22.4 | +01:59:59 | 11.29 | 1724 | 444 | 79.8 | 170 |
| 5792 | Lowell | 3 | 14:57:51.5 | −01:02:29 | 12.08 | 1929 | 415 | 75.5 | 84 |
| 5813 | Lowell | −5 | 15:00:40.2 | +01:45:39 | 11.45 | 1926 | 250 | 43.6 | 145 |
| 5850 | Lowell | 3 | 15:06:39.4 | +01:36:13 | 11.54 | 2553 | 255 | 29.4 | 140 |
| 5985 | Lowell | 3 | 15:39:24.1 | +59:22:00 | 11.87 | 2519 | 329 | 57.5 | 13 |
| 6015 | Lowell | 6 | 15:50:51.5 | +62:20:17 | 11.69 | 824 | 322 | 66.5 | 28 |
| 6118 | Lowell | 6 | 16:19:12.6 | −02:09:57 | 12.42 | 1572 | 280 | 64.7 | 58 |
| 6384 | Lowell | 4 | 17:29:59.0 | +07:05:43 | 11.14 | 1667 | 369 | 48.6 | 30 |
| 6503 | Lowell | 6 | 17:48:53.7 | +70:09:00 | 10.91 | 42 | 424 | 70.2 | 123 |

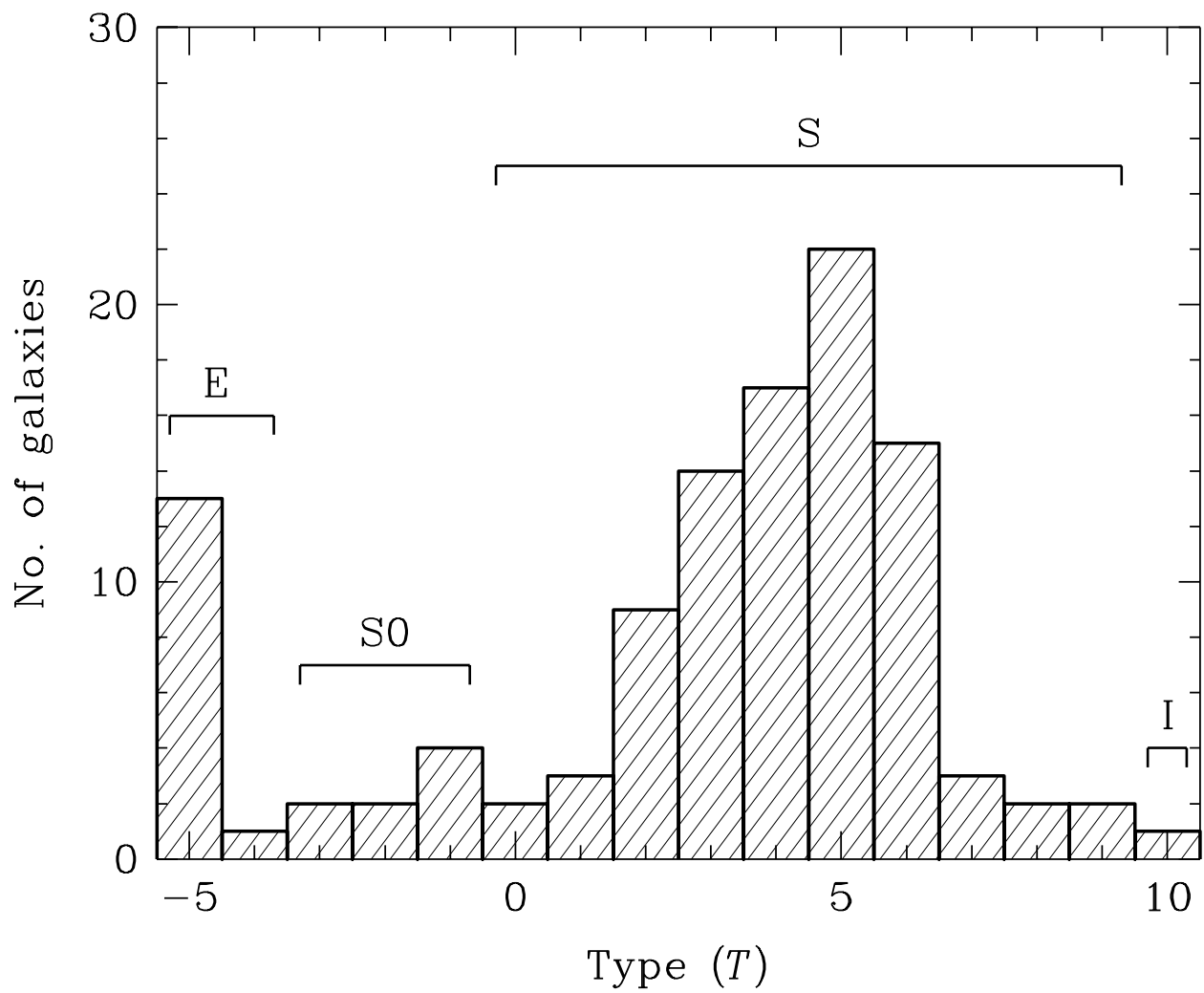

Figure 2.

TABLE 2. FITS header keys.

| Name | Type | Description |
| --- | --- | --- |
| EXPOSURE | float | Length of exposure in seconds. |
| FILTER1 | character | Photometric passband used for the observation. |
| RA | character | Right ascension of the pointing center during the observation. |
| DEC | character | Declination of the pointing center during the observation. |
| AIRMASS | float | Airmass at the beginning of the exposure. |
| TIME | character | Universal time (UT) at the begining of the exposure. |
| SKY | float | The background sky level (in ADU). |
| SKYSIG | float | RMS noise per pixel of the sky (in ADU). |
| DNAT0_ST | float | Counts (in ADU) representing 0 magnitude in the image, calibrated using observations of standard stars. |
| DNAT0_BV | float | Counts (in ADU) representing 0 magnitude in the image, calibrated using $B$ and $B-V$ from RC3. |
| G_CENT_X | integer | Horizontal (x) coordinate of the photometric center of the galaxy in the trimmed image, in pixels. |
| G_CENT_Y | integer | Vertical (y) coordinate of the photometric center of the galaxy in the trimmed image, in pixels. |
| SATURAT | integer | Equals 1 if the galaxy center was not saturated, 2 if it was saturated (and divided by 2). |
| PSF_FWHM | float | Width (in arcsec) of the point spread function, obtained from stars before foreground star removal. |
| B_RC3 | float | Total (asymptotic) Johnson $B$ magnitude from RC3, column 6, line 1. |
| B-V_RC3 | float | Total (asymptotic) Johnson $B-V$ color from RC3, column 7, line 1. |
| PA_RC3 | float | Major axis position angle, (measured from North eastward, between 0° and 180°) from RC3, column 5, line 1. |
| BOA_RC3 | float | Decimal logarithm of the ratio of the major axis isophotal diameter to the minor isophotal diameter, both reduced to $\mu_B = 25$ mag arcsec$^{-2}$, from RC3, column 4, line 2. |
| SIZE_RC3 | float | Decimal logarithm of the apparent major axis isophotal diameter reduced to $\mu_B = 25$ mag arcsec$^{-2}$, from RC3, column 4, line 1. |
| TYPE_RC3 | character | Numerical morphological type $T$ from RC3, column 3, line 1. |
| VELO_RC3 | float | Weighted mean heliocentric radial velocity, calculated from RC3, column 10, line 4. |

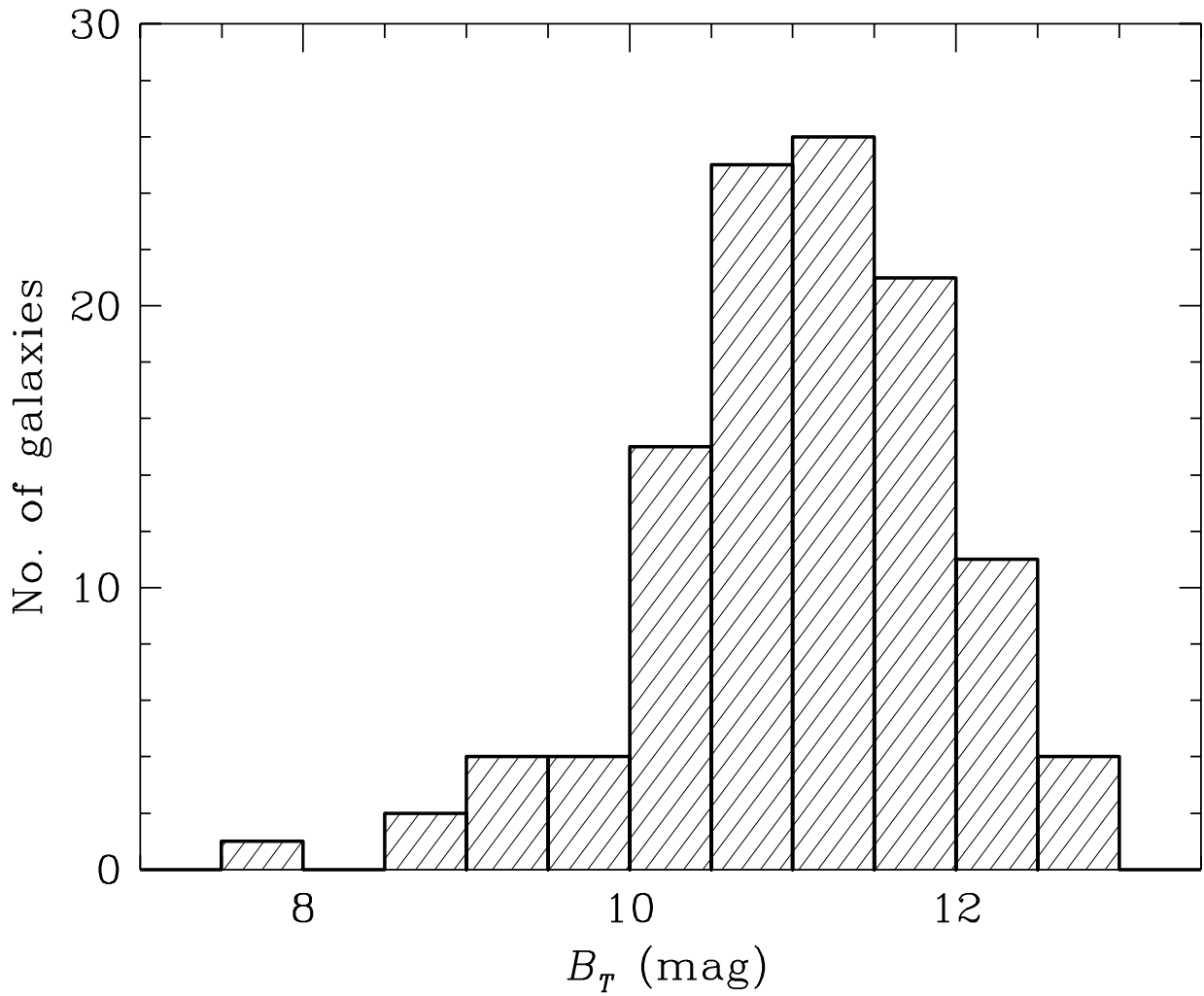

Figure 3.

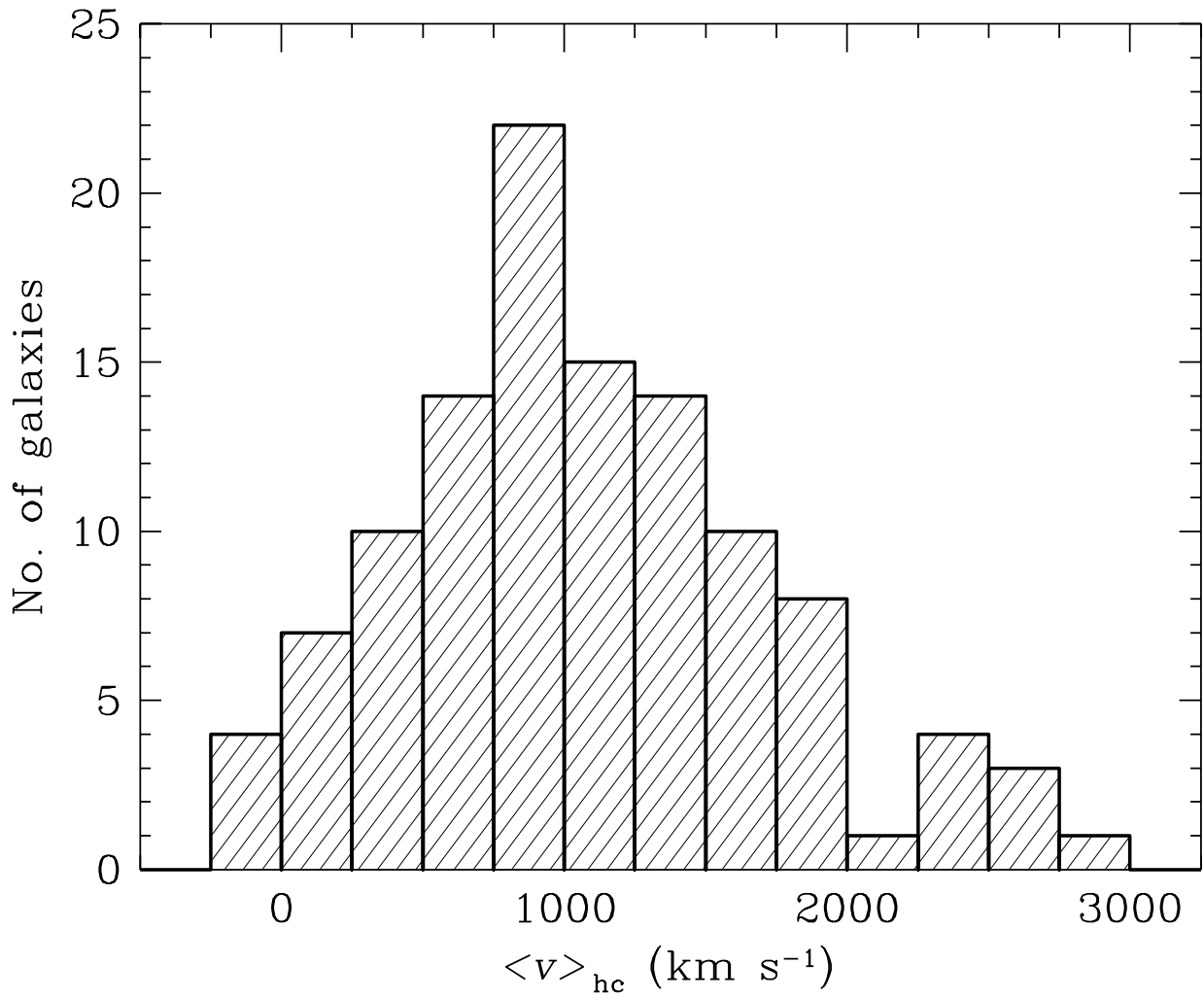

Figure 4.

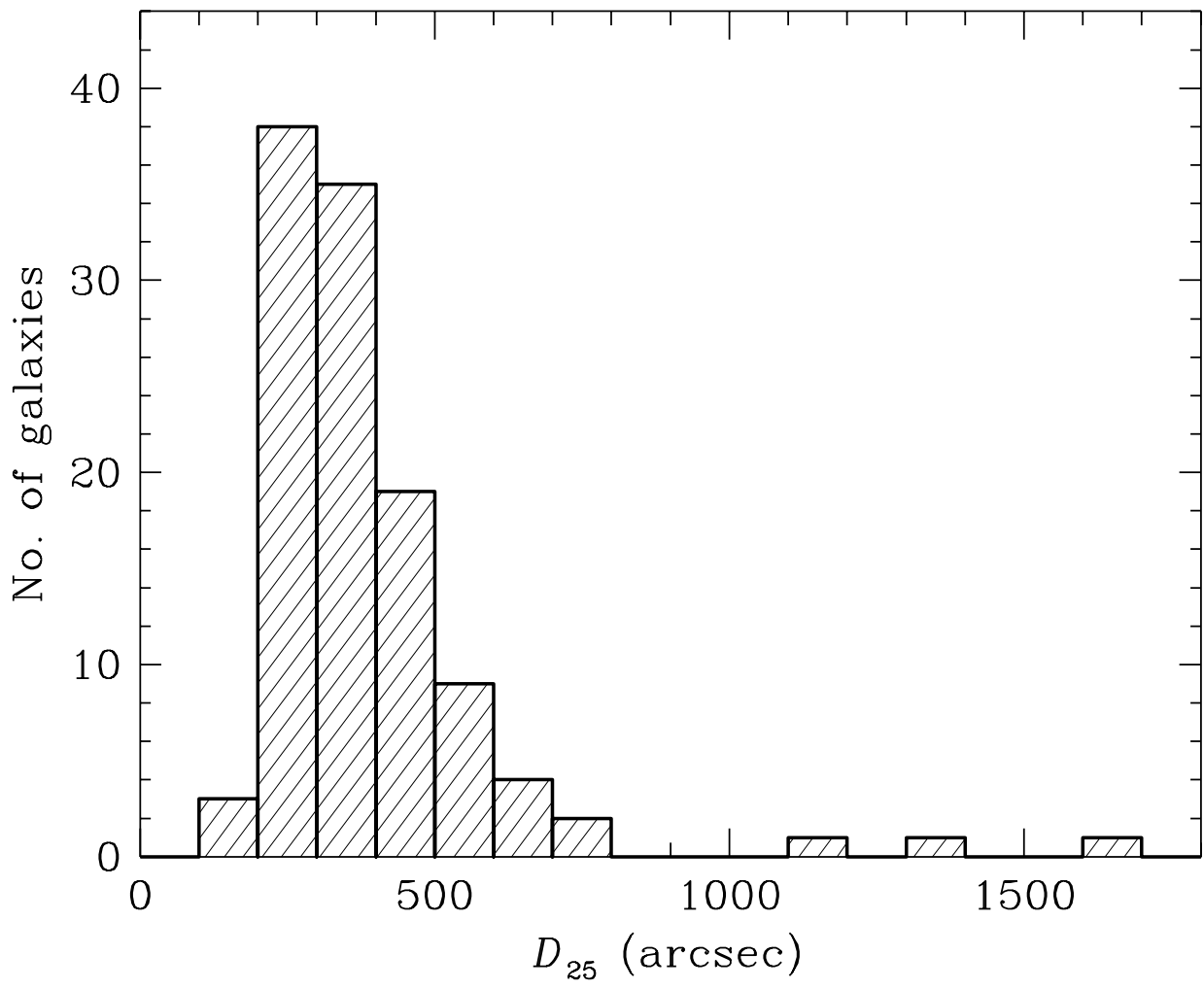

Figure 5.

# A CATALOG OF DIGITAL IMAGES
# OF 113 NEARBY GALAXIES

Zsolt Frei,

Puragra Guhathakurta[1]

and

James E. Gunn

Department of Astrophysical Sciences

Princeton University, Peyton Hall

Princeton, NJ 08544

and

J. Anthony Tyson

AT&T Bell Laboratories

Murray Hill, NJ 07974

[1] Current address: UCO/Lick Observatory, University of California, Santa Cruz, CA 95064

**ABSTRACT**

We present a digital catalog of images of 113 galaxies in this paper. These galaxies are all nearby, bright, large and well resolved. All images were recorded with charge coupled devices (CCDs) at the Palomar Observatory with the 1.5 meter telescope and at the Lowell Observatory with the 1.1 meter telescope. At Palomar we used the Thuan–Gunn $g$, $r$ and $i$ photometric bands (Thuan and Gunn, 1976) to take 3 images each of 31 spiral galaxies; at Lowell we used the $B_J$ and $R$ bands (2 images per galaxy) of the photometric system by Gullixson *et al.* (1994) to observe 82 spirals and ellipticals. The galaxies were selected to span the Hubble classification classes. All data are photometrically calibrated with foreground stars removed. Important data on these galaxies published in the "Third Reference Catalog of Bright Galaxies" (de Vaucouleurs *et al.*, 1991) are recorded in the FITS file headers. All files are available through anonymous FTP from "astro.princeton.edu", and Princeton University Press will soon publish the data on CD-ROM.

## 1. INTRODUCTION

We describe a digital catalog of galaxy images. The direct motivation for the development of such a catalog was the need for a suitable data set to test the automatic galaxy classification techniques we are currently working on (Frei, 1995a, in preparation). Automatic means of classifying galaxies are necessary to handle the huge amounts of imaging data that we soon to be available from large survey projects, such as the Sloan Digital Sky Survey (Gunn and Knapp, 1993). We think, however, that these images are useful beyond the scope of this project; consequently, we make the digital images available. The catalog contains FITS files (Wells *et al.*, 1981) of the images, with an extensive set of corresponding information in the file headers.

To test the classification techniques, we needed a set of galaxies that span the full range of the Hubble morphological classes. Another requirement was to have large, well resolved galaxies so that we could develop our techniques step–by–step: first, solve the easiest challenges that large galaxies pose, then improve the techniques so that they work for galaxies which are further away. The performance and limitations of the classification system can be tested on degraded images. If images of galaxies in the digital catalog are available in several passbands then one can interpolate or extrapolate in color, thus synthetically "redshifting" the image of a given galaxy to any desired $z$ (while decreasing the resolution at the same time).

Consequently, we developed a catalog of images of nearby, well resolved galaxies that were recorded in several passbands with charge coupled devices (CCDs). Although CCDs have been in extensive use for about 15 years, there is no similar data set publicly available which we know of.

Early photometric surveys of galaxies date back to the 70's. Freeman (1970) collected surface photometric data for 36 spiral and lenticular galaxies to examine the disks. Kormendy (1977a, 1977b, 1977c) used photoelectric photometry to study a total of 20 red and neutral compact galaxies. King (1978) performed photographic surface photometry on 17 galaxies, mostly giant ellipticals, to compare the brightness profiles of giant ellipticals to those of star clusters in his famous study (King, 1962, 1965, 1966). Burstein (1979a, 1979b) investigated 18 S0 galaxies. He also used photographic surface photometry to explore the luminosity profiles and the origin of S0 galaxies. Boroson (1981) derived luminosity profiles from photographic photometry of 26 spiral galaxies.

Kent (1984, 1985) did CCD surface photometry of field galaxies. His set contains 105 galaxies of different, widely distributed morphological types. These are not the largest galaxies on the sky; in fact, we found only three common between galaxies in his set and our set of 113 nearby galaxies. His aim in observing field galaxies was, among other things, to compare statistics of field galaxies to statistics of disk galaxies in clusters. Using the photometric data he calculated bulge–to–disk ratios based on intensity profiles along the major and minor axes.

Other recent works on nearby galaxies with CCDs include Davis et al. (1985), who studied three well known elliptical galaxies in detail. Lauer (1985) had a larger sample of 42 galaxies, ellipticals and S0s, which allowed him to do systematic studies of central structures and core properties. Simien and de Vaucouleurs (1986) compiled a large data set to investigate bulge–to–disk ratios of spirals and lenticulars. Their data included 98 galaxies, 32 of which were observed by them. Other data came from works mentioned above (Boroson, 1981; Burstein, 1979b; Kormendy, 1977b) and from two others (Yoshizawa and Wakamatsu, 1975; Whitmore and Kirshner, 1981); not all of these images were taken with CCDs.

Watanabe *et al.* (1982), Watanabe (1983), Okamura *et al.* (1984) and Watanabe *et al.* (1985) studied more than 200 galaxies to develop a quantitative classification system. They used plates and digitized them with an isophotometer which used a one dimensional CCD as a detector. The culmination of their effort is the "Photometric Atlas of Northern Bright Galaxies" (Kodaira *et al.*, 1990). This book contains digitized images of 791 galaxies. Unfortunately, these data are available in a printed form only, and since the images are digitized from photographic plates, all problems presented by those plates (nonlinearity, for example) are incorporated into the digital form of the data. More recently, Han (1992) presented CCD surface photometry of spirals (284 galaxies in 16 clusters) and Colless *et al.* (1993) studied ellipticals and lenticulars (352 galaxies in distant clusters). However, the galaxies in the last two studies are too distant for the purposes of developing and testing our automated classification system.

There are other catalogs with images of galaxies, all of which are printed from photographic plates (Sandage, 1961; Sandage and Bedke, 1988). Also worth mentioning are those efforts that resulted in a large collection of useful photometric data on galaxies, but which do not contain images (Sandage and Tammann, 1981; Tully and Fisher, 1987; de Vaucouleurs *et al.*, 1991).

Images of the galaxies in our catalog were obtained at two different sites. At the Lowell Observatory, we used the 1.1 meter telescope and a CCD camera. The photometric system is described in Gullixson *et al.* (1994). We used two filters, $B_J$ and $R$. About 25% of our images were observed at the Palomar Observatory with the 1.5 meter telescope equipped with a CCD camera and the Thuan-Gunn photometric system (Thuan and Gunn, 1976). We used three passbands: $g$, $r$, and $i$.

An unusual feature of our catalog is that foreground stars have been removed from the frames. This has been done for several reasons: to derive quantitative parameters that describe the galaxy for the purposes of classification, we had to ensure that the foreground objects did not contribute to these parameters. Secondly, to synthetically "redshift" the images of these galaxies, the images had to be free of foreground stars. We think that star–removal will also be useful for other possible applications of this catalog.

In the next section, we describe the observations. Section 3 contains details of the processing of the images. In Section 4, we tabulate and plot distributions of parameters of the galaxies in the catalog, and in Section 5 we describe the image file format and the information contained in the file headers. In the short discussion in Section 6, we give details on how to access the data through the computer network.

## 2. OBSERVATIONS

### 2.1 *Palomar*

Thirty-one galaxies were observed with the 1.5 meter telescope (P60) of Palomar Observatory on the night of 1991 May 4 with the Wide Field Prime Focus Universal Extragalactic Instrument (Wide Field PFUEI). Gunn and Westphal (1981) describe the original PFUEI, designed for the 5 meter telescope (P200) at Palomar. The wide field camera has a 306 mm f/4 aerial camera lens collimator and a Nikon 58 mm f/1.2 camera lens. The field of view is $16' \times 16'$, and it is projected on an $800 \times 800$ Texas Instruments (TI) CCD in the Cassegrain focus, yielding a scale of $1''.19$ pixel$^{-1}$. The focal ratio of the system is f/1.65 (for more details see Hester and Kulkarni, 1989).

Images were obtained in the $g$, $r$ and $i$ bands of the Thuan–Gunn photometric system. The original, 4–band system ($u$, $v$, $g$ and $r$) was introduced by Thuan and Gunn (1976). It was later extended by Wade *et al.* (1979) with the $i$ band, and by Schneider *et al.* (1983) with the $z$ band. Most CCDs are not efficient in the $z$ band, but $g$, $r$ and $i$ CCD surface photometry of galaxies have been extensively done in the past decade. These filters are centered at 500, 650 and 820 nm, respectively. See Frei and Gunn (1994) for a detailed description of the photometric system we used, including the system response curves (including the effects of atmospheric absorption, two reflections from aluminum, the transmission functions of filters and optics, and the quantum efficiency of the detector). The 7.5 cm square filters were placed in front of the camera optics to avoid problems with reflection.

All exposures were 60 s. The 9 electron CCD readout noise was low enough to allow such a short exposure time and still ensure that the background was sky noise limited. The gain ($G$) of the instrument is about 2 electrons/ADU (analog–to–digital units). The effective noise in electrons ($\sigma_e$) is the quadratic sum of the readout noise and the sky noise (the latter, ideally, being the square root of the signal in electrons in a given pixel):

$$\sigma_e = \sqrt{R^2 + NG} \quad , \tag{1}$$

where $R$ is the readout noise in electrons and $N$ is the counts in ADU in a given pixel. With 60 s exposure time, the level of the sky was typically around 900, 1300 and 2200 counts in the $g$, $r$ and $i$ bands, respectively. Using (1) it is obvious that the readout noise is negligible, and signals at the sky level are background noise limited. Typical counts around the center of galaxies are 10000–20000 ADU, making the readout noise even less significant there. The sky noise that is actually measured off the galaxy in the images is slightly higher than $\sigma_e$ suggested by (1) due to imperfections in flat fielding and the presence of faint objects and cosmic ray events in the region where the noise is measured.

Keeping the exposure time short for all the observations made it possible to obtain more than 150 images (including flats and standard star observations) in one night. The centers of some of our brightest galaxies, however, are saturated. Saturation was limited to only few pixels around the photometric center of these galaxies. We introduced a scheme to repair these regions and to calibrate these galaxies (see the next section).

### 2.2 *Lowell*

Eighty two galaxies were observed with the 1.1 meter telescope of the Lowell Observatory on several nights between 1989 March 24 and April 4. A camera with a thinned, back–illuminated 320 × 512 pixel RCA CCD was in the Cassegrain focus.

A 2:1 focal reducer gave an f/4 focal ratio, yielding a scale of $1''.35$ pixel$^{-1}$. The field of view is about $7' \times 11'$. For more details, see Guhathakurta and Tyson (1989).

We used the three–band $B_J$, $R$ and $I$ photometric system, which was introduced to take advantage of silicon CCDs (Gullixson et al., 1994). Only $B_J$ and $R$ images were obtained of each galaxy. These filters are centered at 450 and 650 nm. Frei and Gunn (1994) give system response curves representing the system as we used it.

Typical exposure times were between 100 and 600 s, depending on the galaxy core brightness. This yielded sky levels between 100 and 1000 counts (the sky in the $R$ band being typically twice as high as in the $B_J$ band for identical exposure times). Short text exposures were first obtained of each galaxy in each filter in order to avoid core saturation. The readout noise of the CCD was 90 electron, and the gain of the instrument was 11 electron/ADU. From (1) we conclude that in the background the readout noise and the sky noise are about equally important. Typical counts in central regions of the galaxies reach above 10000 ADU, so there the shot noise dominates over the readout noise.

Both observing runs took place in spring. Since we tried to avoid galaxies close to the Galactic equator, all the galaxies are located in the Northern Galactic cap. The positions of the galaxies in the catalog are plotted in Figure 1 in equatorial coordinates. We used a Hammer–Aitoff projection which produces an equal area map of the entire sky. The Galactic plane is shown with a thick curve for orientation. The CCD images are oriented such that North is up, and East is to the left.

## 3. Data Processing

Debiasing and flat fielding the images of galaxies observed at Palomar was straightforward. We used the data in the extended horizontal register of the CCD to obtain the bias level, which was subtracted from the image. Twilight sky flats were obtained in all three passbands during the night of observations. These were used to create a median flat which was used to flat field the images.

Besides subtracting the bias level, we had to use a more complicated flattening procedure for the images observed at Lowell. Because of the glass substrate of the RCA CCD, the night sky emission lines caused Fabry–Perot fringing in the $R$ band ($B_J$ is not affected since there are no strong atmospheric emission lines in this band). Because of fringing, we treated the $R$ and the $B_J$ images differently. Another problem was that observations took place in both bright and dark conditions, since the observing run was longer than a week. The color of the background sky is close to the color of the twilight flat in bright conditions, but can be quite different if the Moon is down. Consequently, we handled images taken in bright and dark conditions differently. Some of these techniques were used by Tyson (1986).

For the $B_J$ images, if the background was bright, we divided by a twilight flat. Because the colors are similar, this was sufficient to obtain good results. If the sky was dark, after dividing by the twilight flat we also subtracted a carefully scaled twilight flat to account for the color difference. For the $R$ frames, where the fringing amplitude can be as large as 15% of the night sky level, in addition to the steps performed for the $B_J$ frames, we also subtracted a scaled "fringe frame". Twilight flats have a much lower fringe amplitude than even the bright sky images.

We scaled the fringe frame to remove fringes as best we could from the twilight flat, then used it to flat–field the object exposures. We constructed the fringe frame by imaging relatively sparse parts of the sky for 300 s at least five times each night, and by averaging these frames by median filtering to minimize contributions from individual stars and galaxies.

3.1 *Core Saturation in Palomar Images*

A few pixels around the photometric centers of some of the galaxies observed at Palomar were saturated. Compared to the total area of the fields, only a very small portion of the images were affected, accounting for only a negligible part of the total light of the galaxy. We modeled the light distribution around the photometric centers – $L(x, y)$ – with a seven–parameter, two dimensional Gaussian distribution of the form:

$$L(x,y) = A + B \exp\left[-C_1(x-a)^2 - C_2(y-b)^2 - C_3(x-a)(y-b)\right] \quad , \quad (2)$$

where $a$ and $b$ are the coordinates of the photometric center of the given galaxy, $x$ and $y$ are the rectangular coordinates in the image, $A$, $B$, $C_1$, $C_2$, and $C_3$ are constants. These 5 constants and the coordinates of the center of the galaxy are to be fitted.

First, to preserve integer representation of the images, we divided all pixel values by 2 in the entire image if the center was saturated; in no case did this compromise the sampling of the noise histogram. Next, we used the information in the image just around the saturated center to fit these parameters, and then replaced all saturated pixels with analytical values from (2). Since the few images which are saturated were overexposed only slightly, dividing by 2 was enough in all cases to ensure that the fitted centers are lower than the maximum value, 32767, that can be represented by short signed integers in the pixels. We made a clear indication whether the images were saturated (and fixed by this method) in the file headers (see Section 5).

The fitting algorithm was the multidimensional downhill simplex method of Nelder and Mead (1965). The actual code is adopted from Press *et al.*'s (1988) *amoeba* routine. Crucial for the success of the fit is the selection of sensible starting values of the parameters to be fitted. The estimated center of the galaxy ($a$ and $b$) was obtained as a user input; the background sky level ($A$) and the cross–term ($C_3$) were set to 0. The widths in $x$ and $y$ dimensions ($C_1$ and $C_2$) were estimated from the full width at half maximum (FWHM) of the point spread function (PSF) representing the image. The peak value ($B$) was set twice as high as the saturation level (32767). With these starting values the iteration converged quickly and gave good results.

Eleven of the 31 spiral galaxies observed at Palomar were saturated at the photometric center, eight of them in all bands, three of them only in two bands. All saturated images were fitted with the above method and their saturated pixels replaced. Different extents of saturation in the different passbands resulted in the Gaussian distribution being fitted to different regions of the same galaxy. Because of this, colors may be unreliable in the very central regions of these galaxies.

### 3.2 *Star Removal*

We used an empirical two dimensional PSF to fit and remove the foreground stars from the images. The program we developed for this purpose first identifies stars off the galaxy to be used for constructing the PSF. Second, it finds those objects which are likely to be foreground stars (versus HII regions, bright stars belonging to the galaxy, etc) and removes them using a PSF fit. Third, the residuals left in the image are repaired cosmetically. Although the program does not perform all necessary steps automatically, it was very useful for the purpose of cleaning these images. Details of this procedure are to be found in a separate paper (Frei, 1995b).

On several occasions the area occupied by the galaxy was much smaller than the total area of the original image, and we chose to extract square regions out of the original images that contained the galaxy, which saved a considerable amount of storage space. The CCD chip used at Lowell was not square, making it necessary to trim to obtain square images. We decided to construct the catalog using only square images of a few standard sizes. Offsets (the coordinates of the lower left corner of the trimmed image in the original image) and the coordinates of the photometric center of the galaxies in the final, clipped images were recorded in the file headers (see Section 5).

We also calibrated all the galaxies photometrically, but not without difficulty. Conditions during several nights of observations at Lowell, and part of the observing night at Palomar were non–photometric. Moreover, observations of standard stars were not available for all nights at Lowell. In a separate paper (Frei and Gunn, 1994), we derived transformation relations for colors of galaxies among five different photometric systems. These were used to calculate magnitudes in the given filter systems from data available in the Third Reference Catalog of Bright Galaxies (de Vaucouleurs *et al.*, 1991, RC3). For those data from Palomar, both the calculated magnitudes and the magnitudes obtained from standard–star–calibrations are given in the file headers (there were seven standard stars observed during the night of observations). The Lowell data contain only magnitudes calculated via color transformations of the RC3 data.

We summed up total counts corresponding to the galaxy in a circle centered on the photometric center of the galaxy (the $x$ and $y$ coordinates of which is recorded in the file header). The radius of this circle was half of the horizontal (or vertical, since all images are square) size of the image (which is also recorded in the file header). This aperture is large enough to effectively make our magnitude the total magnitude of the galaxy. We subtracted the contribution of the sky using the total area of the circle and the sky level recorded in the file header. If the photometric center of the galaxy does not coincide with the center of the image, we used only the part of the circle which is inside the image for the sum, and calculated the area of this part for correct sky subtraction. With all the necessary information available from the FITS header keys, these steps can be reproduced with ease. Starting from this total count representing the galaxy, and the calibrated magnitude in the given band (derived via color transformation of RC3 data), we obtained the count which would represent 0 magnitude in a pixel; this is stored in the file header.

— 15 —## 4. Properties of Galaxies in the Catalog

All of the galaxies in our catalog are listed in RC3. We recorded some of the properties found in RC3 for these galaxies in the file headers (such as morphological type, total $B$ magnitude, $B - V$ color, etc). Here we present distributions of some of the most important properties, to statistically describe the catalog. In the next section, we will give detailed information about all the parameters included in the file headers. Although we developed techniques — some of them new — to derive similar properties and to automatically process the images for classification, the aim of this paper is to report the data, not our reduction technique. Thus we chose to present parameters from RC3, a well known source. We will present the properties we obtained with our methods for these galaxies elsewhere (Frei, 1995a, in preparation).

The revised morphological type on the "Handbuch der Physik" system is used in RC3 (de Vaucouleurs, 1959, 1963; de Vaucouleurs *et al.*, 1991). The histogram of the numerical type $T$ of the galaxies in our catalog is in Fig. 2. Among the 113 galaxies, one is of type I0 (non–Magellanic irregular). The numerical type ($T$) is 90 for this galaxy in RC3. We omitted this galaxy from the histogram. The distribution of morphological types is well suited to our classification task. We have many more spirals than ellipticals. Deriving parameters of spirals (based on the fine structure in the disk) is more complicated than finding representative parameters of ellipticals (where fine structure, although well studied, is not usually considered for classification purposes).

Total $B_T$ magnitudes are derived from photoelectric aperture photometry and from surface photometry in RC3. The histogram in Fig. 3 shows the distribution of this parameter. Although we calibrated our images in their respective bands, there is not a single band in which observations are available for all the galaxies (we used different photometric systems at Palomar and at Lowell). Consequently, the easiest way to compare brightnesses is to use data from RC3. As a rough measure of the distances of the galaxies, we plotted the histogram of the radial velocities corrected to the reference frame of the cosmic microwave background radiation ($v_{3K}$) in Fig. 4. Isophotal major axis diameters ($D_{25}$) listed in RC3 were obtained from photographic and CCD surface photometry and from photoelectric growth curves. We produced a histogram of these sizes in Fig. 5, where the diameters are expressed in arcseconds. The majority of our galaxies are 4–6 arcmin in diameter, some of them even larger; our catalog contains some of the brightest, largest, most nearby galaxies around the Northern Galactic cap.

RC3 lists isophotal axis ratios for the galaxies, both ellipticals and spirals. We calculated inclination angles (I) from these data according to

$$I = \cos^{-1}\left(\frac{1}{10^{R_{25}}}\right) \quad , \tag{3}$$

where $R_{25}$ is the decimal logarithm of the axis ratio (as it is actually given in RC3, obtained at an isophotal surface brightness level $\mu_B = 25$ mag arcsec$^{-2}$). As usual, 0° represents face–on galaxies, while 90° corresponds to edge–on orientation. We plotted a histogram of the inclination angles for our spiral and lenticular galaxies (99 in total) in Fig. 6, since the axis isophotal ratio is not directly related to an inclination angle for ellipticals. The apparent lack of face–on galaxies is understandable since spiral galaxies, even if face–on, will have non-circular isophotes at any brightness level due to the uneven surface brightness distribution of light in the disk (spiral arms). Edge–on, or almost edge–on galaxies will have isophotes which appear "fatter" than the orientation of the disk would indicate, due to finite disk thickness and due to the presence of the bulge. This will result in the measured inclination angle (from (3)) being somewhat (systematically) larger than the "true" disk inclination.

In Fig. 7 we plot the distribution of position angles. The standard convention for measuring position angles is from the North eastward between 0° and 180°, as it is given in RC3. It is reassuring that they are evenly distributed (there is no unreasonable selection effect in our catalog). The apparent excess of galaxies in the first bin (0 position angle) is because the position angle was artificially set to 0 for any nearly face–on spirals in RC3 (for those with inclination angles less than $\approx 40°$).

Some of the most important properties of the galaxies are given in Table 1. The first column is the NGC number of the catalog galaxy. The observatory where the images were obtained is identified in column 2. All Palomar galaxies are in three bands (three separate images are available, in $g$, $r$ and $i$), while all Lowell galaxies are in two bands (two images, in $B_J$ and $R$). The numerical revised morphological type $T$ from RC3 is in column 3; these data are used to construct the histogram in Fig. 2. Columns 4 and 5 are right ascensions and declinations as recorded in the file headers at the time of observations. Columns 6 to 10 are the total $B_T$ magnitudes, radial velocities corrected to the reference frame of the microwave background radiation ($v_{3K}$), major axis isophotal diameters ($D_{25}$), inclination angles, and position angles, respectively. These data are all from RC3, and they were used for Figs. 3–7.

## 5. FITS header keys

We recorded useful information in the header keys in the FITS files of the images. There are several keys among those we used which are *standard* FITS keys (for the description of the FITS format and the standard keys see Wells *et al.*, 1981; or "Definition of the Flexible Image Transport System (FITS)", from NOST). The standard keys that we populated with data are: SIMPLE, BITPIX, NAXIS, NAXIS1, NAXIS2, OBJECT, TELESCOP, EPOCH, BSCALE, BZERO, BUNIT, DATAMAX, DATAMIN and DATE-OBS.

There are several keys that we added to describe the data as best as we could. All necessary information is in Table 2. The first column contains the name of the header key, the second column has the type of the variable, and the last column gives a short description. The first six keys (EXPOSURE to TIME) were recorded during the observations. We obtained SKY and SKYSIG from the data at the time of calibration (the sky level is the median of all pixels in the image). The photometric zeropoints DNAT0_ST and DNAT0_BV are calculated from observations of standard stars (not available for the Lowell data) and by using data from RC3, respectively. We recorded the offsets of the trimmed, smaller image in the original, full size image and the photometric center of the galaxy in the trimmed, square image. The parameter SATURATE is used to tell whether the galaxy center was saturated and repaired. The last seven header keys are populated with data from RC3. These are well described in RC3 and in the previous section. We gave the location (column and line) of the actual data entry in the RC3 table.

## 6. Conclusion

During 1989–1990, a substantial set of CCD surface photometry of nearby, well resolved galaxies were collected from Palomar and Lowell. The data were used to construct a catalog of images in FITS format, containing a uniform set of information in the file headers. The catalog is suitable for our ongoing galaxy classification project (Frei, 1995a, in preparation), and we believe it can also be used for other purposes. We collected and presented information about the observations and various data processing steps in the previous sections. The description of FITS header keys and brief statistics of key properties of the catalog galaxies are also included to aid the prospective user of these data.

To our knowledge, no data similar to that presented here are available to the general public today. Although we plan to increase the number of galaxies in the catalog, we think that the current set of images is already useful for various applications, and thus it should be made accessible to anyone interested. The image files are on our "anonymous" FTP site at "astro.princeton.edu". The data are in "frei/Galaxies". Additional information is in the README file in the same directory. The data may also be accessed via World Wide Web (WWW) at URL "ftp://astro.princeton.edu/frei/Galaxies".

## 7. Acknowledgements

It is our pleasure to acknowledge the great number of people who participated in the observations and data reductions in the past 5 years. Their work made it possible to construct the catalog as presented here (we obviously try to avoid saying "to complete", since such a large set of data can be improved and extended almost indefinitely). Students at the Department of Astrophysical Sciences, Princeton University, did observations at Lowell as part of a student project. We thank Jill Knapp for organizing the project. We are indebted to several people at Lowell: Jay Gallagher, who made it possible to use the observatory for the project, Craig Gullixson, who helped with the telescope, Neal Hartsough (University of Arizona) and Rick Wenk (AT&T Bell Labs) who built the software system. We are greatly indebted to the former students who took part in the data acquisition and reduction: Arif Babul, Renyue Cen, Charles Gammie, Bin Gao, Lynne Hillenbrand, Neil Katz, Man Hoi Lee, Joanna Lees, Kevin Long, Shude Mao, Jordi Miralda–Escude, Changbom Park, Duncan Walsh and David Weinberg. We acknowledge very useful discussions with Robert Lupton and Neil Tyson, who also helped to revise the manuscript. The patience of the Palomar time allocation committee, who continued to give time for this project through five completely clouded–out years, is also much appreciated.

This research was supported in part by NSF through grant no. AST–9100121. Z. F. acknowledges the support of OTKA through grant no. F4491.

– 22 –

Kent, S. M. 1985, ApJ, 59, 115

King, I. R. 1962, AJ, 67, 471

King, I. R. 1965, AJ, 70, 376

King, I. R. 1966, AJ, 71, 64

King, I. R. 1978, ApJ, 222, 1

Kodaira, K., Okamura, S., & Ichikawa, S. 1990, Photometric Atlas of Northern Bright Galaxies (University of Tokyo Press, Tokyo)

Kormendy, J. 1977a, ApJ, 214, 359

Kormendy, J. 1977b, ApJ, 217, 406

Kormendy, J. 1977c, ApJ, 218, 333

Lauer, T. R. 1985, ApJS, 57, 473

Nelder, J. A., & Mead, R. 1965, Computer Journal, 7, 308

NOST: NASA/Science Office of Standards and Technology 1993, Definition of the Flexible Image Transport System (FITS), Code 633.2 (NASA Goddard Space Flight Center, Greenbelt, MD)

Okamura, S., Kodaira, K., & Watanabe, M. 1984, ApJ, 280, 7

Press, H. W., Flannery, B. P., Teukolsky, S. A., & Vetterling, W. T. 1988, Numerical Recipes in C (Cambridge University Press, Cambridge)

Sandage, A. 1961, The Hubble Atlas of Galaxies (Carnegie Institution of Washington, Washington, D. C.)

Sandage, A., & Bedke, J. 1988, Atlas of Galaxies (NASA, Washington, D. C.)

Sandage, A., & Tammann, G. A. 1981, A Revised Shapley–Ames Catalog of Bright Galaxies (Carnegie Institution of Washington, Washington, D. C.)

Schneider, D. P., Gunn, J. E., & Hoessel, J. G. 1983, ApJ, 264, 337

Simien, F., & de Vaucouleurs, G. 1986, ApJ, 302, 564

Thuan, T. X., & Gunn, J. E. 1976, PASP, 88, 543

– 24 –## FIGURE CAPTIONS

**Figure 1.** Positions of the 113 galaxies in the catalog on the sky. An equal–area Hammer–Aitoff projection was used. The center is at $12h$ RA and $0°$ declination. The Galactic plane is shown with the thick curve. All galaxies are around the Northern Galactic cap, and avoid the Galactic plane.

**Figure 2.** Distribution of the numerical morphological types $T$ of the galaxies in the catalog. $T$ is obtained from RC3 for all the galaxies. Types $-5$ and $-4$ are ellipticals, $-3$ to $-1$ are lenticulars, 0 to 9 are spirals and the rest are irregulars.

**Figure 3.** Distribution of the total $B_T$ magnitudes of the galaxies in the catalog. $B_T$ is from RC3 for all the galaxies.

**Figure 4.** Distribution of the velocities, corrected to the reference frame of the cosmic microwave background radiation, of the galaxies in the catalog. $v_{3K}$ is from RC3 for all the galaxies.

**Figure 5.** Major axis isophotal diameters of the galaxies in the catalog. These diameters, at a surface brightness level $\mu_B = 25$ mag arcsec$^{-2}$, are found in RC3 for all the galaxies.

**Figure 6.** Distribution of inclination angles of the *spiral, lenticular and irregular* galaxies in the catalog. The data are found in RC3 for all the galaxies. The apparent lack of face–on and edge–on galaxies are due to the techniques used to determine the inclination angles. See Section 4 of the text for an explanation.

**Figure 7.** Distribution of position angles of all galaxies in the catalog. All data are from RC3. The high number of galaxies in the first bin is due to the fact that all nearly face–on spiral galaxies were assigned 0 position angle in RC3.

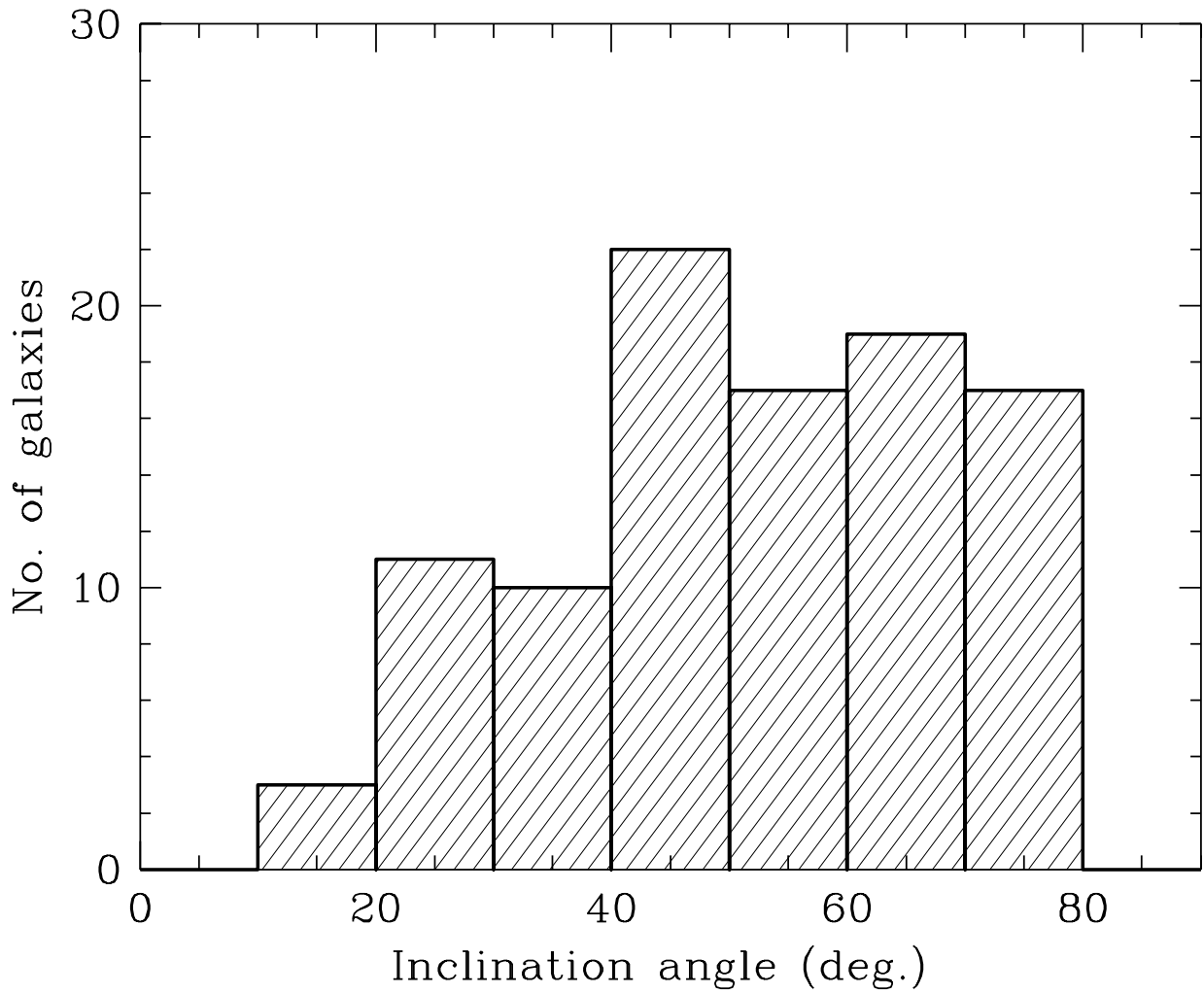

Figure 6.

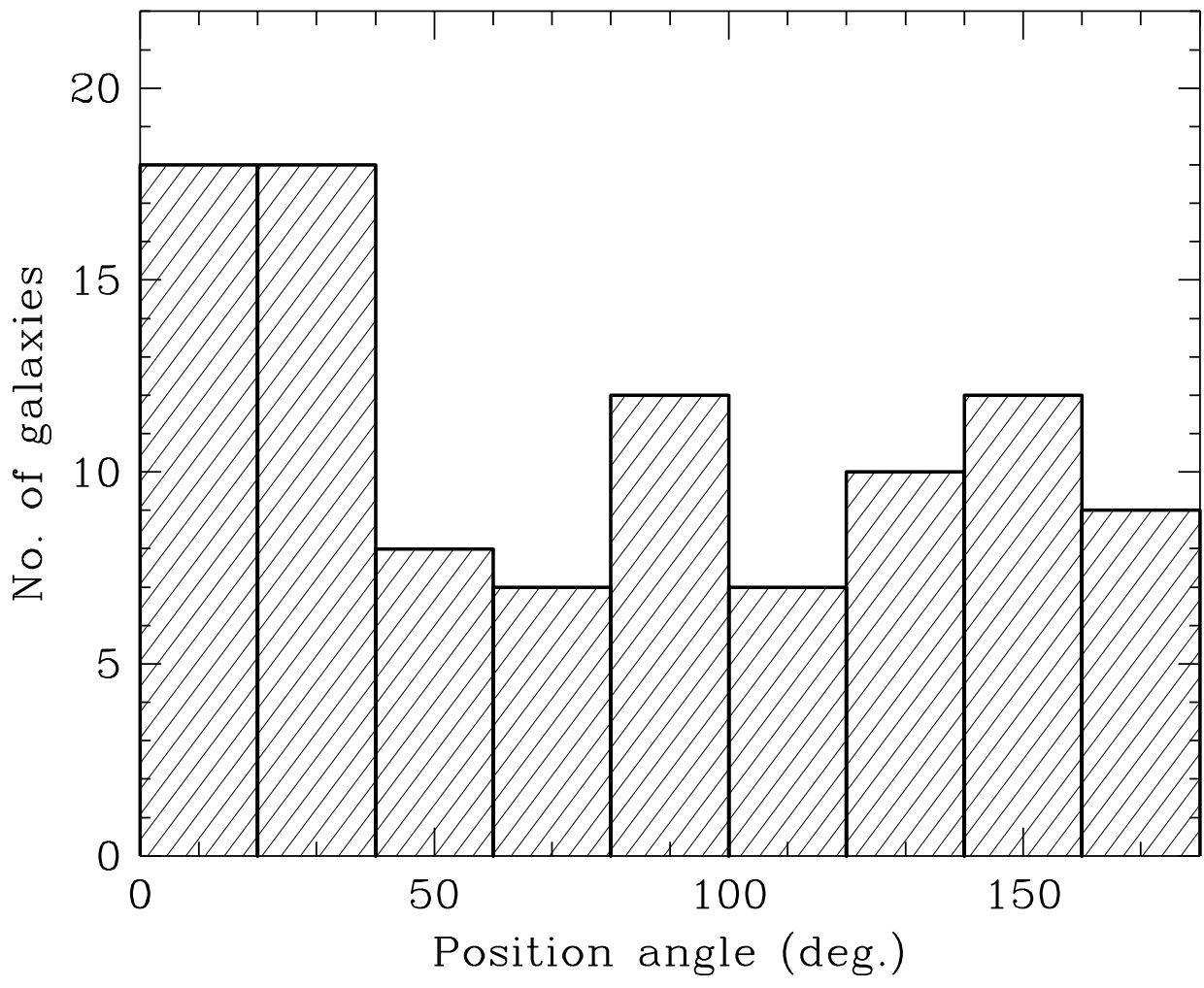

Figure 7.